\documentclass[10pt,twocolumn,english,jcp,aip,english,superscriptaddress]{revtex4-1}
\usepackage[T1]{fontenc}
\usepackage[latin9]{inputenc}
\usepackage{xcolor}
\usepackage{pdfcolmk}
\usepackage{float}
\usepackage{units}
\usepackage{amsmath}
\usepackage{amssymb}
\usepackage{cancel}
\usepackage{graphicx}
\usepackage{esint}
\PassOptionsToPackage{normalem}{ulem}
\usepackage{ulem}

\makeatletter

\newcommand{\lyxdot}{.}

\providecolor{lyxadded}{rgb}{0,0,1}
\providecolor{lyxdeleted}{rgb}{1,0,0}

\@ifundefined{textcolor}{}
{%
 \definecolor{BLACK}{gray}{0}
 \definecolor{WHITE}{gray}{1}
 \definecolor{RED}{rgb}{1,0,0}
 \definecolor{GREEN}{rgb}{0,1,0}
 \definecolor{BLUE}{rgb}{0,0,1}
 \definecolor{CYAN}{cmyk}{1,0,0,0}
 \definecolor{MAGENTA}{cmyk}{0,1,0,0}
 \definecolor{YELLOW}{cmyk}{0,0,1,0}
}

\@ifundefined{textcolor}{}{%
 \definecolor{BLACK}{gray}{0}
 \definecolor{WHITE}{gray}{1}
 \definecolor{RED}{rgb}{1,0,0}
 \definecolor{GREEN}{rgb}{0,1,0}
 \definecolor{BLUE}{rgb}{0,0,1}
 \definecolor{CYAN}{cmyk}{1,0,0,0}
 \definecolor{MAGENTA}{cmyk}{0,1,0,0}
 \definecolor{YELLOW}{cmyk}{0,0,1,0}
}

\usepackage{dcolumn}
\usepackage{bm}
\usepackage{enumerate}

\usepackage{babel}

\usepackage{babel}

\usepackage{babel}

\usepackage{babel}

\makeatother

\usepackage{babel}
\begin{document}

\title{Exact Calculation of the Time Convolutionless Master Equation Generator:
Application to the Nonequilibrium Resonant Level Model }

\author{Lyran Kidon}

\affiliation{School of Chemistry, The Sackler Faculty of Exact Sciences, Tel Aviv
University,Tel Aviv 69978,Israel}

\affiliation{The Sackler Center for Computational Molecular and Materials Science,
Tel Aviv University, Tel Aviv, Israel 69978}

\author{Eli Y. Wilner}

\affiliation{School of Physics and Astronomy, The Sackler Faculty of Exact Sciences,
Tel Aviv University,Tel Aviv 69978, Israel}

\altaffiliation{Current address: Department of Physics, Columbia University, New York, NY 10027, USA.}

\author{Eran Rabani}

\affiliation{Department of Chemistry, University of California and Materials Science
Division, Lawrence Berkeley National Laboratory, Berkeley, California
94720, USA}

\affiliation{The Sackler Center for Computational Molecular and Materials Science,
Tel Aviv University, Tel Aviv, Israel 69978}
\begin{abstract}
The generalized quantum master equation provides a powerful tool to
describe the dynamics in quantum impurity models driven away from
equilibrium. Two complementary approaches, one based on Nakajima--Zwanzig--Mori
time-convolution (TC) and the other on the Tokuyama--Mori time-convolutionless
(TCL) formulations provide a starting point to describe the time-evolution
of the reduced density matrix. A key in both approaches is to obtain
the so called ``memory kernel'' or ``generator'', going beyond
second or fourth order perturbation techniques. While numerically
converged techniques are available for the TC memory kernel, the canonical
approach to obtain the TCL generator is based on inverting a super-operator
in the \emph{full} Hilbert space, which is difficult to perform and
thus, all applications of the TCL approach rely on a perturbative
scheme of some sort. Here, the TCL generator is expressed using a
reduced system propagator which can be obtained from system observables
alone and requires the calculation of super-operators and their inverse
in the \emph{reduced }Hilbert space rather than the full one. This
makes the formulation amenable to quantum impurity solvers or to diagrammatic
techniques, such as the nonequilibrium Green's function. We implement
the TCL approach for the resonant level model driven away from equilibrium
and compare the time scales for the decay of the generator with that
of the memory kernel in the TC approach. Furthermore, the effects
of temperature, source-drain bias, and gate potential on the TCL/TC
generators are discussed.
\end{abstract}
\maketitle

\section{Introduction}

The development of accurate and efficient schemes to calculate the
dynamic response in quantum impurity models remains one of the greatest
challenge in chemical and condensed matter physics. Significant advances
have been made in the context of the spin-boson model,\cite{leggett1987dynamics}
in which a single impurity spin is coupled to a bosonic bath at equilibrium.\cite{Mak1992,Egger1994,Makri1995,Golosov1999,Thoss03,Makri2015}
Addressing the dynamics in quantum impurity models driven away from
equilibrium by the application of a bias voltage, is much more challenging,
and most bruit-force real-time techniques are limited to relatively
short times.\cite{muhlbacher_real-time_2008,Wang2009,werner_diagrammatic_2009,werner_weak-coupling_2010,eckel_comparative_2010,gull10_bold_monte_carlo,Segal10,Huetzen12,Simine13,cohen2014green} 

This short-coming associated with the well-known dynamical sign problem
can be lifted by combining real-time impurity solvers with a reduced-density
matrix formalism.\cite{cohen_memory_2011,wilner_bistability_2013}
The basic idea is that the total system can be decomposed into two
parts - an interesting part we call ``the system'' and an environment
collectively called ``the bath'' to which the quantum system is
coupled. The exact calculation of the dynamics within this formulation
is often given in terms of the Nakajima--Zwanzig--Mori \emph{time-convolution}
(TC) approach~\cite{nakajima_quantum_1958,zwanzig_ensemble_1960,mori_transport_1965}
derived from the unitary dynamics of the full Hilbert space using
projection operator techniques. In this formalism, the complexity
of solving the many-body quantum Liouville Von-Newmann equation is
reduced to the evaluation of a super-operator so-called the ``memory
kernel'' which fully determines the non-Markovian dynamics of the
system.\cite{Breuer2002} Since the decay of the system typically
exceeds the characteristic decay time of the memory kernel, brute-force
numerical impurity solvers limited to short times can be used to extract
the kernel, from which the long-time dynamics can be obtained.\cite{cohen_memory_2011,cohen_generalized_2013}
This approach has proven fruitful for spin magnetization dynamics~\cite{Cohen2013kondo}
in the nonequilibrium Anderson impurity model~\cite{Anderson1961}
driven by electron-electron correlations and for nonadiabatic dynamics
and bistability~\cite{wilner_bistability_2013,wilner2014,wilner_phonon_2014}
in the nonequilibrium extended Holstein model~\cite{Holstein1959}
driven by electron-phonon correlations. This formalism is also widely
used as a starting point to derive perturbative schemes for fermionic
systems,~\cite{Leijnse2008,Royer2003335} and has gained attention
in other fields as well.\cite{zhang_nonequilibrium_2006,Berkelbach2012,Kelly2013} 

An alternative to the TC generalized quantum master equation is based
on Tokuyama--Mori \emph{time-convolutionless }(TCL) quantum master
equation.\cite{Tokuyama1976statistical} The technique eliminates
the dependence of the time evolution on the history of the system
and replaces the integro-differential TC equation with a first order
differential equation, which is exact and local in time.\cite{Shibata1977,Chaturvedi1979,Breuer1999Pstochastic,Timm2008,Timm2011Pertubative}
In the TCL quantum master equation, the TC memory kernel is replaced
by a time-local super-operator referred to as the TCL generator or
kernel. A direct calculation of the TCL generator is extremely difficult
since it relies on finding an inverse to a super-operator in the full
Hilbert space.\cite{Breuer2002} Thus, in nearly all applications
of the TCL approach, a perturbative scheme in the system-bath coupling
is used to obtain the TCL generator.\cite{Laird1991,Reichman1996,Golosov2001,Timm2011Pertubative,Clos2012,ElsiMari2012,Fischer2013}
An attempt to formulate a non-perturbative path-integral approach
to calculate the kernel has been proposed recently.\cite{Nan2009}

In this work we adopt a different approach to calculate the TCL kernel
based on the reduced density propagator~\cite{Laird1991} and show
that the matrix elements of this 4-ranked tensor can be obtained directly
from the population and coherences of the reduced system alone. While
this connection may seem redundant, in fact, it offers two important
advantages: First, since the TCL generator decays on a faster time-scale
compared to the decay of the populations and coherences themselves,
one can use numerically converged impurity solvers to obtain the TCL
generator at short-times and infer about the converged system dynamics
on any time scale, similar to the cutoff method used in the TC approach.\cite{cohen_memory_2011,cohen_generalized_2013}
Second, an alternative diagrammatic expansion based on the nonequilibrium
Green's function (NEGF) approach offers means to systematically expand
the kernel beyond second or fourth order in the system-bath coupling.

We implement the TCL approach to study the population dynamics in
the resonant level model~\cite{Haug2008}, analyze the time scales
for the decay of the TCL kernel as a function of temperature ($T$),
source-drain bias ($V_{{\rm SD}}$) and gate voltage ($\varepsilon$),
and compare it to the decay of the TC memory kernel. We find that
both kernels decay on an identical time scale and thus, in this respect,
neither formulations is advantageous. However, the TCL formulation
requires only system dependent quantities as input, quantities that
are often easier to generate using a numerically converged impurity
solvers.

\section{Reduced density matrix formalism: TC and TCL approaches}

\subsection{The reduced density matrix formalism }

Consider an open quantum system coupled to an environment representing
the ``bath'' degrees of freedom, described by the full Hamiltonian:
\begin{equation}
H=H_{0}+V=H_{S}+H_{B}+V.\label{eq:genertic H}
\end{equation}
where $H_{0}=H_{S}+H_{B}$ describes the system and bath Hamiltonians,
and $V$ the coupling between the two. The time evolution of the full
density matrix is given by the Liouville--von-Neumann equation, $\dot{\rho}\left(t\right)=-\frac{i}{\hbar}\left[H,\rho\left(t\right)\right]\equiv{\cal L}\rho\left(t\right)$,
for which a numerical converged solution is impossible to obtain due
to the exponential scaling complexity with the size of the system
and bath. However, in most situations the dynamics of the bath is
not interesting and thus, one is concerned with the time evolution
of the system alone, described by the reduced density matrix, $\sigma\left(t\right):$
\begin{eqnarray}
\rho_{B}\otimes\sigma\left(t\right)={\cal P}\rho\left(t\right) & \equiv & \rho_{B}\otimes\mathrm{Tr}_{B}\rho\left(t\right),
\end{eqnarray}
where $\rho_{B}$ is the density matrix of the initial state of the
bath and ${\cal P}=\rho_{B}\otimes\mathrm{Tr}_{B}\cdots$ is a projection
operator onto the system subspace. There are two different approaches
to describe the time evolution of $\sigma\left(t\right)$: (a) the
Nakajima--Zwanzig--Mori \emph{time-convolution} approach~\cite{nakajima_quantum_1958,zwanzig_ensemble_1960,mori_transport_1965}
and (b) the Tokuyama--Mori \emph{time-convolutionless} approach.\cite{Tokuyama1976statistical} 

For the former (TC), the equation of motion for $\sigma\left(t\right)$
is given by
\begin{equation}
\frac{\partial}{\partial t}\sigma\left(t\right)=\mathcal{L}_{S}\sigma\left(t\right)+\int_{0}^{t}d\tau\kappa\left(\tau\right)\sigma\left(t-\tau\right)\label{eq:sigma(t)}
\end{equation}
where ${\cal L}_{S}=-\frac{i}{\hbar}\left[H_{S},\cdots\right]$ and
the memory kernel (a super-operator in the system subspace) is given
by
\begin{equation}
\kappa\left(t\right)=Tr_{B}\left\{ \mathcal{L}e^{{\cal Q}\mathcal{L}t}{\cal Q}\mathcal{L}\rho_{B}\right\} \label{eq:memory-kernel}
\end{equation}
with ${\cal P}+{\cal Q}={\cal I}$ (${\cal I}$ is the unit operator).
For the latter (TCL), the equation of motion for the reduced density
operator is given in terms of a time-local kernel:\cite{Breuer2002}
\begin{equation}
\frac{\partial}{\partial t}\sigma\left(t\right)={\cal K}\left(t\right)\sigma\left(t\right).\label{eq:TCL QME}
\end{equation}
In the above equation, the TCL kernel (a super-operator in the system
subspace) is defined by

\begin{align}
{\cal K}\left(t\right) & =\mathrm{Tr}_{B}\left\{ \mathcal{L}\left(1-\Sigma\left(t\right)\right)^{-1}\rho_{B}\right\} \label{eq:TCL Generator}
\end{align}
where 
\begin{eqnarray}
\Sigma\left(t\right) & = & \int_{0}^{t}d\tau e^{{\cal Q}{\cal L}\tau}{\cal Q}\mathcal{L}{\cal P}e^{-{\cal L}\tau}
\end{eqnarray}
is a super-operator in the \emph{full }Hilbert space of the system
and bath. In the above equations, we have assumed a factorized initial
condition, $\rho\left(0\right)=\rho_{B}\left(0\right)\otimes\sigma\left(0\right)$.
For a correlated initial state, an additional term should be included
in both the TC and TCL equations for $\sigma\left(t\right)$.\cite{Breuer2002}

The complication of solving for the reduced density operator is manifested
through the TC or TCL kernels. In fact, it seems more complicated
to solve for the kernels than obtaining the full density operator
$\rho\left(t\right)$, since both $\kappa\left(t\right)$ and ${\cal K}\left(t\right)$
involve projected propagation $e^{{\cal Q}{\cal L}t}$ and furthermore,
the TCL kernel requires an inversion of a super-operator in the full
Hilbert space. However, the reduced density operator formalism offers
two main advantages. First, when formulated in the interaction picture,
the TC and TCL approaches are useful as a starting point for approximate
methods, such as the second order perturbation in the system-bath
coupling leading to the Redfield equations. Second, if the dynamics
governing $\kappa\left(t\right)$ or ${\cal K}\left(t\right)$ are
short lived, then obtaining the dynamics of $\sigma\left(t\right)$
on all time scales can be achieved by calculating the kernels at short
time only. In fact, this has been a useful approach to obtain the
memory kernel in the Nakajima-Zwanzig-Mori TC approach, as recently
illustrated for nonequilibrium impurity models~\cite{cohen_memory_2011,Cohen2013kondo,wilner_bistability_2013,wilner_phonon_2014,wilner2014}
and for other condensed phase systems.\cite{Berkelbach2012,Kelly2013}

\subsection{The TCL Generator in terms of the reduced system propagator}

In the Nakajima-Zwanzig-Mori TC approach, one can rewrite the memory
kernel in terms of a Volterra equation of the second kind, removing
the complexity of the projected dynamics of Eq.~(\ref{eq:memory-kernel}):\cite{zhang_nonequilibrium_2006}
\begin{equation}
\kappa\left(t\right)=\frac{\partial\Phi\left(t\right)}{\partial t}-\Phi\left(t\right)\mathcal{L}_{S}-\int_{0}^{t}d\tau\Phi\left(t-\tau\right)\kappa\left(\tau\right)\label{eq:volterra}
\end{equation}
where $\Phi\left(t\right)=Tr_{B}\left\{ \mathcal{L}_{V}e^{\mathcal{L}t}\rho_{B}\right\} $
and ${\cal L}_{V}=-\frac{i}{\hbar}\left[V,\cdots\right]$. The super-operator
$\Phi\left(t\right)$ can now be calculated by a variety of numerically
exact techniques,\cite{weiss_iterative_2008,muhlbacher_real-time_2008,Wang2009,gull10_bold_monte_carlo,Segal10}
since it does not involve projected dynamics. A similar approach for
the TCL generator seems difficult to derive, since one has to invert
the super-operator $1-\Sigma\left(t\right)$ which spans the entire
Hilbert space. Indeed, most applications of the TCL approach were
based on expanding $\left(1-\Sigma\left(t\right)\right)^{-1}$ in
a Taylor series with the condition that $\left|\Sigma\left(t\right)\right|\le1$,
such that ${\cal K}\left(t\right)$ is obtained perturbatively.\cite{Breuer2002} 

Here, we adopt a simple formalism to obtain ${\cal K}\left(t\right)$
circumventing the need to invert a super-operator in the full Hilbert
space by rewriting the TCL generator in terms of system observables
only. The approach is thus, amenable to impurity solvers of the kind
used in the TC approach.\cite{weiss_iterative_2008,muhlbacher_real-time_2008,Wang2009,gull10_bold_monte_carlo,Segal10}
We begin by redefining the expression for $\sigma\left(t\right)$
as~\cite{Laird1991}
\begin{eqnarray}
\sigma\left(t\right) & = & \mathrm{Tr}_{B}\left\{ \rho\left(t\right)\right\} =\mathrm{Tr}_{B}\left\{ {\cal U}\left(t\right)\rho\left(0\right)\right\} 
\end{eqnarray}
where ${\cal U}\left(t\right)=e^{\mathcal{L}t}$ is the full propagator.
For factorized initial conditions,
\begin{eqnarray}
\sigma\left(t\right) & = & \mathrm{Tr}_{B}\left\{ {\cal U}\left(t\right)\sigma\left(0\right)\otimes\rho_{B}\left(0\right)\right\} \nonumber \\
 & = & \mathrm{Tr}_{B}\left\{ e^{iHt}\sigma\left(0\right)\otimes\rho_{B}\left(0\right)e^{-iHt}\right\} \nonumber \\
 & \equiv & {\cal U}_{S}\left(t\right)\sigma\left(0\right)\label{eq:sigma(t)=00003DUsigma(t)sigma(0)}
\end{eqnarray}
where ${\cal U}_{S}\left(t\right)$ is the propagator of the system
(dot) only (${\cal U}_{S}\left(t\right)\neq e^{{\cal L}_{S}t}$).
By reversing the equation and performing a time derivative on it,
we obtain~\cite{Laird1991}
\begin{equation}
\dot{\sigma}\left(t\right)=\dot{{\cal U}}_{S}\left(t\right)\sigma\left(0\right)=\dot{{\cal U}}_{S}\left(t\right){\cal U}_{S}^{-1}\left(t\right)\sigma\left(t\right).
\end{equation}
This equation has the form of the TCL quantum master equation for
the reduced density matrix (Eq.\ \eqref{eq:TCL QME}) where by analogy
the TCL generator ${\cal K}\left(t\right)$ is given by:\cite{Laird1991}
\begin{equation}
{\cal K}\left(t\right)=\dot{{\cal U}}_{S}\left(t\right){\cal U}_{S}^{-1}\left(t\right).\label{eq:K=00003DdU*U-1}
\end{equation}
Thus, in order to obtain the TCL generator one has to compute ${\cal U}_{S}\left(t\right)$
as defined in Eq.~\eqref{eq:sigma(t)=00003DUsigma(t)sigma(0)} and
also invert ${\cal U}_{S}\left(t\right)$. As will become clear below,
obtaining the matrix elements of ${\cal U}_{S}\left(t\right)$ is
straightforward and so is the inversion. To see this, we rewrite the
above equations in a basis set and provide explicit expression for
the elements of the reduced propagator in the following subsection.

\subsection{Matrix representation\label{sub:Matrix-Representation}}

Choosing a basis for the system subspace as the eigenstates of the
system decoupled from the bath, $\left|i\right\rangle $, an operator
$O$ will be represented by its matrix elements $O_{ij}=\left\langle i\left|O\right|j\right\rangle $
and a super-operators $\mathcal{O}$ will be represented by the elements
of a ``tetradic'' (specified by four subscripts) $\mathcal{O}_{ij,kl}=\mathrm{Tr}_{S}\left\{ \left(\left|i\right\rangle \left\langle j\right|\right)^{\dagger}\mathcal{O}\left|k\right\rangle \left\langle l\right|\right\} .$
A super-operator is an operator that works on other operators rather
than on quantum states and thus, it turns a matrix into a new matrix
in the following way:
\begin{equation}
\left({\cal O}O\right)_{ij}=\sum_{kl}\mathcal{O}_{ij,kl}O_{kl}\label{eq:SuperOperator on Operator}
\end{equation}
Following this definition, it can be shown that the product of two
super-operators is given by:
\begin{equation}
\left(\mathcal{OR}\right)_{ij,kl}=\sum_{mn}O_{ij,mn}\mathcal{R}_{mn,kl}.\label{eq:Super-Operator * Super-Operator}
\end{equation}
Therefore, the time evolution of the reduced density matrix elements
is given by:
\begin{equation}
\sigma_{ij}\left(t\right)=\sum_{kl}{\cal U}_{S,ij,kl}\left(t\right)\sigma_{kl}\left(0\right),\label{eq:sigma_ij(t)=00003DU_ijkl(t)sigma_kl(0)}
\end{equation}
the time-local master equation for the reduced density matrix in matrix
form reads:
\begin{equation}
\frac{\partial}{\partial t}\sigma_{ij}=\sum_{kl}{\cal K}_{ij,kl}\left(t\right)\sigma_{kl}\left(t\right),
\end{equation}
and the TCL generator elements are given by:
\begin{equation}
{\cal K}_{ij,kl}\left(t\right)=\sum_{mn}\left({\cal \dot{U}}_{S}\left(t\right)\right)_{ij,mn}\left({\cal U}_{S}^{-1}\left(t\right)\right)_{mn,kl}.\label{eq:Kijkl=00003DdUijmn*U^-1mnkl}
\end{equation}
For simplicity, instead of writing operators in the system sub-space
as $N\times N$ matrices, where $N$ is the dimension of the sub-space,
and super-operators as tensors of dimensions $N^{4}$, we may represent
the first by vectors of size $N^{2}$ and the latter by $N^{2}\times N^{2}$
matrices. This representation preserves the definitions in Eqs.\textbf{~}\eqref{eq:SuperOperator on Operator}
and \eqref{eq:Super-Operator * Super-Operator} and simplifies certain
operations, such as finding the inverse of ${\cal U}_{S}$.

\subsection{System propagator matrix elements\label{sub:System-Propagator-Matrix}}

The exact form of the system propagator matrix elements can be written
explicitly in terms of the matrix elements of the full propagator,
as derived in Appendix~\ref{apx:Tetradic-representation-of-Usigma}.
The derivation is completely general and therefore, also constitutes
a proof that such a tetradic representation (Eq.~\eqref{eq:sigma_ij(t)=00003DU_ijkl(t)sigma_kl(0)})
in fact exists. However, this tetradic form given in the Appendix
may be difficult to treat. 

Here, instead we represent the matrix elements of ${\cal U}_{S}\left(t\right)$
in terms of the reduced density matrix elements directly. This is
done by propagating the system from different initial conditions of
$\sigma\left(0\right)$. For example, if we take $\sigma_{mm}\left(0\right)=1$,
where the subscript $m$ denotes one of the many-body states of the
system, and the remaining values of $\sigma_{ij}\left(0\right)=0,$
then from Eq.~\ref{eq:sigma_ij(t)=00003DU_ijkl(t)sigma_kl(0)} it
follows that 
\begin{equation}
{\cal U}_{S,ij,mm}\left(t\right)=\sigma_{ij}\left(t\right).\label{eq:U diag elements from sigma}
\end{equation}
Hence, the elements ${\cal U}_{S,ij,mm}\left(t\right)$ can be obtained
directly from the matrix elements of the reduced density operator
with the above specified initial condition. Since $\sigma_{ij}\left(t\right)$
can be generated using a proper impurity solver at short times,\cite{weiss_iterative_2008,muhlbacher_real-time_2008,Wang2009,gull10_bold_monte_carlo,Segal10}
the TCL kernel can be generated without the need to invert a super-operator
in the full Hilbert space or use a perturbation expansion to obtain
${\cal K}\left(t\right).$ Rather, the TCL kernel is generated by
inverting ${\cal U}_{S}\left(t\right)$ in the system subspace only.
This amounts to a computational complexity that is similar to that
used to generate the memory kernel in the TC approach.\cite{cohen_memory_2011}

Similarly, for the elements ${\cal U}_{S,ij,mn}\left(t\right)$ with
$m\neq n$, one has to take two different initial conditions for the
system. The first is $\sigma_{mm}\left(0\right)=1$,$\sigma_{mn}\left(0\right)=\sigma_{nm}^{*}\left(0\right)=z$
and the remaining values of $\sigma_{ij}\left(0\right)=0$ and the
second only differs in $\sigma_{mn}\left(0\right)=\sigma_{nm}^{*}\left(0\right)=z'$,
where $z$ and $z'$ are complex numbers ($z\ne z'$) and at least
one is not purely real or imaginary. With the above initial conditions,
we find that:\\
\begin{align}
\sigma_{ij}^{z}\left(t\right) & ={\cal U}_{S,ij,mm}\left(t\right)+z{\cal U}_{S,ij,mn}\left(t\right)+z^{*}{\cal U}_{S,ij,nm}\left(t\right)\nonumber \\
\sigma_{ij}^{z'}\left(t\right) & ={\cal U}_{S,ij,mm}\left(t\right)+z'{\cal U}_{S,ij,mn}\left(t\right)+z'^{*}{\cal U}_{S,ij,nm}\left(t\right).\label{eq:U non-diag elements from sigma}
\end{align}
where the superscript $z$ or $z'$ indicates which initial condition
is used to generate $\sigma\left(t\right)$. Extracting ${\cal U}_{S,ij,mn}\left(t\right)$
and ${\cal U}_{S,ij,nm}\left(t\right)$ can now be achieved by solving
the above linear equations. As before, $\sigma_{ij}^{z}\left(t\right)$
and $\sigma_{ij}^{z'}\left(t\right)$ can be generated at short times
from an impurity solver.\cite{weiss_iterative_2008,muhlbacher_real-time_2008,Wang2009,gull10_bold_monte_carlo,Segal10}

\section{Explicit formulation of the TCL approach for the resonant level model}

\subsection{Model and factorized initial conditions}

We now turn to demonstrate the approach outlined above to calculate
the TCL kernel and compare the time scales governing it with the TC
memory formalism for a transport model system. We focus on the noninteracting
resonant level model since the TCL and TC kernels can be obtained
in terms of an exact nonequilibrium Green's function approach. However,
more complicated models can be treated in the same manner with the
approach presented above. The Hamiltonian describing this open quantum
system is already given in the format of Eq.\ \eqref{eq:genertic H},
where
\begin{equation}
H_{S}=\varepsilon d^{\dagger}d
\end{equation}
is the system Hamiltonian with fermionic creation/annihilation operators
$d^{\dagger}/d$, respectively, and energy $\varepsilon$. The bath
Hamiltonian represents the noninteracting leads:
\begin{equation}
H_{B}=H_{L}+H_{R}=\sum_{k\in L}\varepsilon_{k}c_{k}^{\dagger}c_{k}+\sum_{k\in R}\varepsilon_{k}c_{k}^{\dagger}c_{k},
\end{equation}
with fermionic creation/annihilation operators $c_{k}^{\dagger}/c_{k}$,
respectively for the left ($L$) or right ($R$) leads. The hybridization
between the system and the leads is given by 
\begin{equation}
V=\sum_{k\in L,R}t_{k}\left(c_{k}^{\dagger}d+h.c.\right)
\end{equation}
with coupling strength $t_{k}=\sqrt{\delta\varepsilon{\cal J}\left(\varepsilon_{k}\right)/2\pi}$
and $\delta\varepsilon$ is the band discretization width. For all
applications reported below, we assume a wide band spectral function
of the form: 
\begin{equation}
{\cal J}\left(\varepsilon\right)=\frac{\Gamma/2}{\left(1+e^{\gamma\left(\varepsilon-\varepsilon_{C}\right)}\right)\left(1+e^{-\gamma\left(\varepsilon+\varepsilon_{C}\right)}\right)},
\end{equation}
where $\nicefrac{1}{\gamma}=4\Gamma$ is the cutoff length and $\varepsilon_{C}=40\Gamma$
is the band cutoff energy. 

We take a factorized form for the initial condition described by

\[
\rho\left(0\right)=\sigma\left(0\right)\otimes\rho_{L}\left(0\right)\otimes\rho_{R}\left(0\right)
\]
where $\sigma\left(0\right)$ will be specified below and
\[
\rho_{L/R}\left(0\right)=\frac{1}{Z_{L/R}}\exp\left(-\beta\left(H_{L/R}-\mu_{L/R}N_{L/R}\right)\right)
\]
with $Z_{L/R}$ the corresponding normalization, $\beta=1/k_{B}T$
is the inverse temperature, $\mu_{L/R}$ are the left lead and the
right lead chemical potentials, respectively, such that the source-drain
bias is given by $V_{{\rm SD}}=\mu_{L}-\mu_{R}$. In the above, $N_{L/R}=\sum_{k\in L/R}c_{k}^{\dagger}c_{k}$.

\subsection{Calculation of the TCL kernel}

For the above model, the system is spanned by two levels, an empty
($\left|0\right\rangle $) and occupied ($\left|1\right\rangle =d^{\dagger}\left|0\right\rangle $)
dot. Thus, the reduced density matrix of the system, $\sigma\left(t\right)=\mathrm{Tr}_{B}\left\{ \rho\left(t\right)\right\} $,
is a $2\times2$ matrix and the system propagator ${\cal U}_{S}\left(t\right)$
is a super-matrix with $2^{4}$ elements. For simplicity we will represent
$\sigma\left(t\right)$ with a 4-dimensional vector and ${\cal U}_{S}\left(t\right)$
as a $4\times4$ matrix, as explained in Sec.\ \ref{sub:Matrix-Representation}.
The diagonal matrix elements and coherences dynamics are independent
in the resonant level model, since coherences do not couple different
Fock spaces. Therefore, the system propagator has a block structure
with vanishing off-diagonal block-elements. Since we are interested
in the populations only, we limit the discussion to the diagonal elements
of the reduced density matrix, given by the relevant part of Eq.\ \eqref{eq:sigma(t)=00003DUsigma(t)sigma(0)}:
\begin{align}
\sigma_{00}\left(t\right) & ={\cal U}_{S,00,00}\left(t\right)\sigma_{00}\left(0\right)+{\cal U}_{S,00,11}\left(t\right)\sigma_{11}\left(0\right),\nonumber \\
\sigma_{11}\left(t\right) & ={\cal U}_{S,11,00}\left(t\right)\sigma_{00}\left(0\right)+{\cal U}_{S,11,11}\left(t\right)\sigma_{11}\left(0\right).\label{eq:sigma00=00003DU0000sigma00+U0011sigma11}
\end{align}
Since $\sigma_{11}\left(t\right)=1-\sigma_{00}\left(t\right)$ and
thus ${\cal U}_{S,00,00}\left(t\right)+{\cal U}_{S,11,00}\left(t\right)=1$
and ${\cal U}_{S,00,11}\left(t\right)+{\cal U}_{S,11,11}\left(t\right)=1$,
we can express the relevant part of the reduced propagator in terms
of two matrix elements alone:
\begin{eqnarray}
{\cal U}_{S}\left(t\right) & = & \left(\begin{array}{cc}
{\cal U}_{S,00,00}\left(t\right) & {\cal U}_{S,00,11}\left(t\right)\\
1-{\cal U}_{S,00,00}\left(t\right) & 1-{\cal U}_{S,00,11}\left(t\right)
\end{array}\right).
\end{eqnarray}
From Eq.\ \eqref{eq:sigma00=00003DU0000sigma00+U0011sigma11}, it
is clear that ${\cal U}_{S,00,00}\left(t\right)=\sigma_{00}\left(t\right)$
for an initial empty dot and ${\cal U}_{S,00,11}\left(t\right)=\sigma_{00}\left(t\right)$
for an initial occupied dot. Thus, to obtain the two independent elements
of ${\cal U}_{S}\left(t\right)$, we carry out two separate calculations
corresponding to these two initial dot preparations, where $\sigma_{00}\left(t\right)=\left\langle d\left(t\right)d^{\dagger}\left(t\right)\right\rangle $
can be calculated in a myriad of ways (exactly, for this model). Here,
we use the NEGF approach to obtain $\left\langle d\left(t\right)d^{\dagger}\left(t\right)\right\rangle $.

The TCL kernel matrix elements, ${\cal K}_{00,00}\left(t\right)$
and ${\cal K}_{00,11}\left(t\right)$, can be obtained from Eq.\ \eqref{eq:K=00003DdU*U-1},
and are given by:
\begin{align}
{\cal K}_{00,00}\left(t\right) & =\frac{\dot{{\cal U}}_{S,00,00}\left(1-{\cal U}_{S,00,11}\right)-\dot{{\cal U}}_{S,00,11}\left(1-{\cal U}_{S,00,00}\right)}{{\cal U}_{S,00,00}-{\cal U}_{S,00,11}},\nonumber \\
{\cal K}_{00,11}\left(t\right) & =\frac{\dot{{\cal U}}_{S,00,11}{\cal U}_{S,00,00}-\dot{{\cal U}}_{S,00,00}{\cal U}_{S,00,11}}{{\cal U}_{S,00,00}-{\cal U}_{S,00,11}}.\label{eq:K as func. of U}
\end{align}
The above sum rules for $\sigma\left(t\right)$ and ${\cal U}_{S}\left(t\right)$
translate to the following sum rules for ${\cal K}\left(t\right)$:
\begin{align}
{\cal K}_{00,00}\left(t\right) & =-{\cal K}_{11,00}\left(t\right),\nonumber \\
{\cal K}_{00,11}\left(t\right) & =-{\cal K}_{11,11}\left(t\right).
\end{align}

\subsection{Temperature and Bias Voltage Dependence of ${\cal K}\left(t\right)$}

The temperature and bias voltage dependence of ${\cal K}\left(t\right)$
is complicated, however 
\begin{equation}
{\cal K}_{00,00}\left(t\right)-{\cal K}_{00,11}\left(t\right)=\frac{\dot{{\cal U}}_{S,00,00}\left(t\right)-\dot{{\cal U}}_{S,00,11}\left(t\right)}{{\cal U}_{S,00,00}\left(t\right)-{\cal U}_{S,00,11}\left(t\right)}\label{eq:K-diff}
\end{equation}
is independent of $T$ or $V_{{\rm SD}}.$ This can be shown by using
the approach outlined in Ref.~\onlinecite{Sharma2015} to express
${\cal U}_{S,00,00}\left(t\right)$ and ${\cal U}_{S,00,11}\left(t\right)$.
We begin by defining a correlation matrix with elements $C_{ij}\left(t\right)=\left\langle c_{i}^{\dagger}\left(t\right)c_{j}\left(t\right)\right\rangle $,
where $i,j=0$ refers to the dot ($c_{0}\equiv d$), $i,j=1,\ldots n$
refers to the left lead and $i,j=n+1,\ldots2n$ refers to the right
lead. For a non-correlated initial state, $C\left(0\right)$ is a
diagonal matrix with elements $\left\langle c_{0}^{\dagger}\left(0\right)c_{0}\left(0\right)\right\rangle =n_{d}\equiv f_{0}$,
where $n_{d}$ is the initial population of the dot, $\left\langle c_{i}^{\dagger}\left(0\right)c_{i}\left(0\right)\right\rangle =f_{L}\left(\varepsilon_{i}\right)\equiv f_{i}$
for $1\le i\le n$, and $\left\langle c_{i}^{\dagger}\left(0\right)c_{i}\left(0\right)\right\rangle =f_{R}\left(\varepsilon_{i}\right)\equiv f_{i}$
for $n+1\le i\le2n$. Here, $f_{L/R}\left(\varepsilon\right)=\left(1+e^{\beta\left(\varepsilon-\mu_{L/R}\right)}\right)^{-1}$
is the Fermi Dirac distribution of the left/right lead, respectively.
Next, we define a new set of fermionic operators $a_{\alpha}=\sum_{j=0}^{2n}M_{\alpha j}c_{j}$
where $M$ is the matrix that diagonalizes the one-body Hamiltonian
\begin{equation}
H_{1}=\left(\begin{array}{ccccccc}
\varepsilon & t_{1}^{L} & t_{2}^{L} &  & \cdots &  & t_{n}^{R}\\
t_{1}^{L} & \varepsilon_{1}^{L}\\
t_{2}^{L} &  & \ddots &  &  & 0\\
 &  &  & \varepsilon_{n}^{L}\\
\vdots &  &  &  & \varepsilon_{1}^{R}\\
 &  & 0 &  &  & \ddots\\
t_{n}^{R} &  &  &  &  &  & \varepsilon_{n}^{R}
\end{array}\right)
\end{equation}
with eigenvalues $\tilde{\varepsilon}_{\alpha}=\left(M^{\dagger}\cdot H_{1}\cdot M\right)_{\alpha\alpha}$
and $a_{\alpha}\left(t\right)=e^{-\frac{i}{\hbar}\tilde{\varepsilon}_{\alpha}t}a_{\alpha}\left(0\right)$.
Using these relations, $\sigma_{11}\left(t\right)=1-\sigma_{00}\left(t\right)$
is given by:

\begin{eqnarray}
\sigma_{11}\left(t\right) & = & \left\langle c_{0}^{\dagger}\left(t\right)c_{0}\left(t\right)\right\rangle =\sum_{\alpha\beta}M_{0\alpha}M_{\beta0}^{*}\left\langle a_{\alpha}^{\dagger}\left(t\right)a_{\beta}\left(t\right)\right\rangle \nonumber \\
 & = & \sum_{\alpha\beta}M_{0\alpha}M_{\beta0}^{*}e^{\frac{i}{\hbar}\left(\tilde{\varepsilon}_{\alpha}-\tilde{\varepsilon}_{\beta}\right)t}\left\langle a_{\alpha}^{\dagger}\left(0\right)a_{\beta}\left(0\right)\right\rangle .
\end{eqnarray}
Rewriting the above in terms of the original set of fermionic operators,
we find: 
\begin{align}
\sigma_{11}\left(t\right) & =\sum_{\alpha\beta}M_{0\alpha}M_{0\beta}^{*}e^{\frac{i}{\hbar}\left(\tilde{\varepsilon}_{\alpha}-\tilde{\varepsilon}_{\beta}\right)t}\nonumber \\
 & \ \ \ \cdot\sum_{ij}M_{\alpha i}^{*}M_{j\beta}\left\langle c_{i}^{\dagger}\left(0\right)c_{j}\left(0\right)\right\rangle \nonumber \\
= & \sum_{i=0}^{2n}\, f_{i}\,\left|\sum_{\alpha}M_{0\alpha}^{*}e^{-\frac{i}{\hbar}\tilde{\varepsilon}_{\alpha}t}M_{\alpha i}\right|^{2}.
\end{align}
Using Eq.~\ref{eq:U diag elements from sigma}, we may now recall
that for a choice of an empty initial state, that is for $f_{0}=n_{d}=0$,
${\cal U}_{S,00,00}\left(t\right)=\sigma_{00}\left(t\right)$ and
for an initial occupied dot ${\cal U}_{S,00,11}\left(t\right)=\sigma_{00}\left(t\right)$.
Thus, for $n_{d}=0$: 

\begin{eqnarray}
{\cal U}_{S,00,00}\left(t\right) & = & 1-\sum_{i=\mathbf{1}}^{2n}\, f_{i}\,\left|\sum_{\alpha}M_{0\alpha}^{*}e^{-\frac{i}{\hbar}\tilde{\varepsilon}_{\alpha}t}M_{\alpha i}\right|^{2}
\end{eqnarray}
and for $n_{d}=1$:
\begin{equation}
{\cal U}_{S,00,11}\left(t\right)={\cal U}_{S,00,00}\left(t\right)-\left|\sum_{\alpha}\left|M_{0\alpha}\right|^{2}e^{-\frac{i}{\hbar}\tilde{\varepsilon}_{\alpha}t}\right|^{2}.\label{eq:U0011}
\end{equation}
From the above it is clear that ${\cal U}_{S,00,00}\left(t\right)-{\cal U}_{S,00,11}\left(t\right)=\left|\sum_{\alpha}\left|M_{0\alpha}\right|^{2}e^{-\frac{i}{\hbar}\tilde{\varepsilon}_{\alpha}t}\right|^{2}$,
which is $T$ and $V_{{\rm SD}}$ independent.

\section{Results for the resonant level model}

\begin{figure}[H]
\includegraphics[width=8cm]{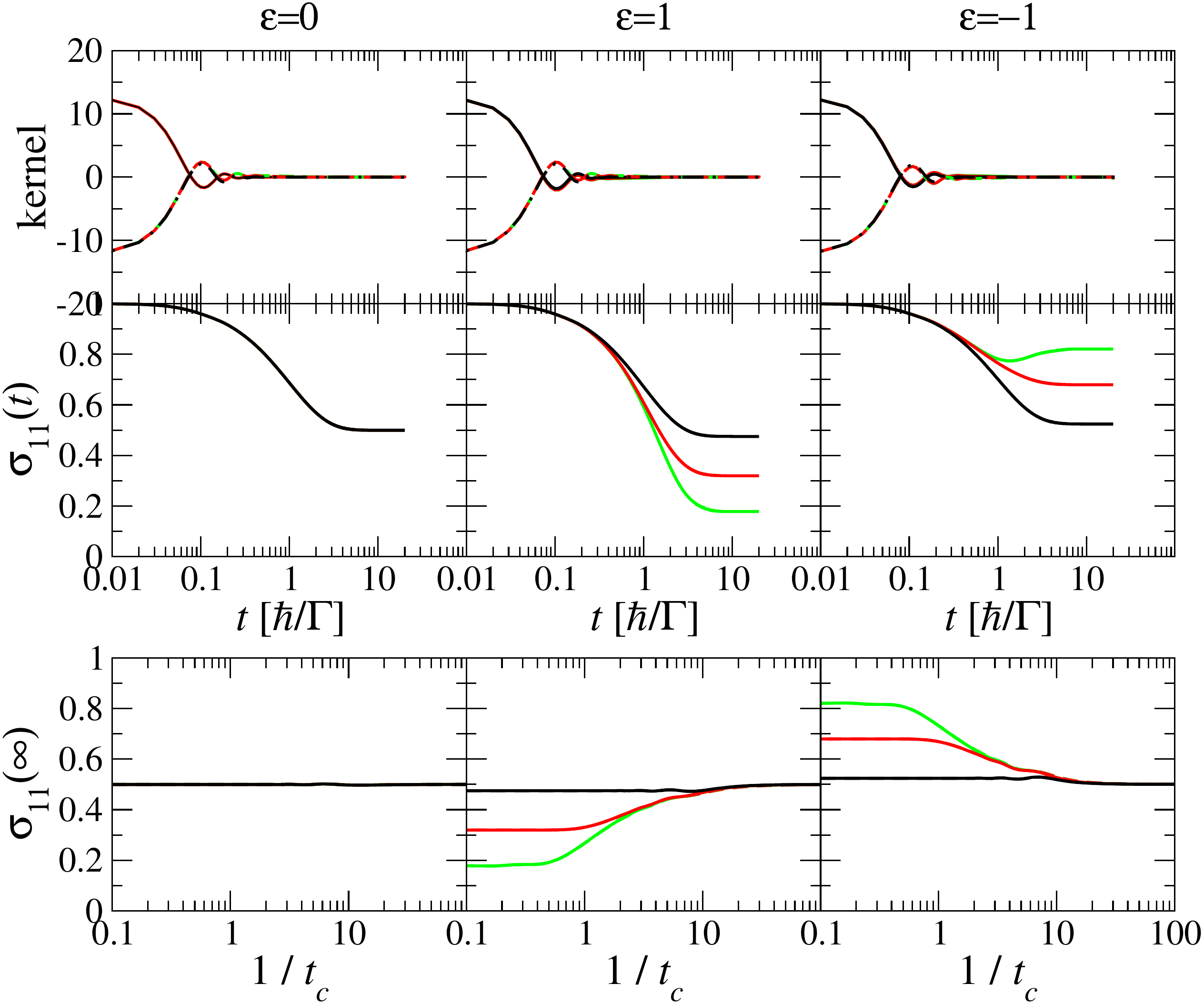}\protect\caption{\label{fig:V=00003D1 different temps} Plots of the steady-state dot
population as a function of the inverse cutoff time (lower panels),
the time dependence of the dot population (middle panels), and $\kappa_{00,00}\left(t\right)$
and ${\cal \partial K}_{00,00}\left(t\right)/\partial t$ (upper panels).
Left, middle and right panels are for different values of the dot
energy, as indicated. The source-drain bias voltage is $V_{{\rm SD}}=\Gamma.$
The green, red and black curves are for $T=\Gamma/10$, $T=\Gamma$,
and $T=10\Gamma$, respectively. }
\end{figure}

To compare the TC and TCL approaches, we have generated the corresponding
kernels for the resonant level model using a two-time NEGF approach,
which is exact for non-interacting electrons. For the resonant level
model, the coherences are decoupled from the populations,\cite{cohen_memory_2011}
and here we focus on the latter only. Hence, both the TC and TCL kernels
have only two independent elements. $\kappa_{00,00}\left(t\right)$
and $\kappa_{00,11}\left(t\right)$ were obtained by solving Eq.~\eqref{eq:volterra},
where the input $\Phi\left(t\right)$ was generated from the two-time
lesser Green's function of the \emph{lead-dot} ($G_{k}^{<}\left(t,t^{\prime}\right)=\frac{i}{\hbar}\left\langle d^{\dagger}\left(t^{\prime}\right)c_{k}\left(t\right)\right\rangle $).
${\cal K}_{00,00}\left(t\right)$ and ${\cal K}_{00,11}\left(t\right)$
were obtained from Eqs.~\eqref{eq:U diag elements from sigma} and
\eqref{eq:K as func. of U}, where the dot population was generated
from \emph{dot} lesser Green's function ($G^{<}\left(t,t^{\prime}\right)=\frac{i}{\hbar}\left\langle d^{\dagger}\left(t^{\prime}\right)d\left(t\right)\right\rangle $). 

\begin{figure}[t]
\includegraphics[width=8cm]{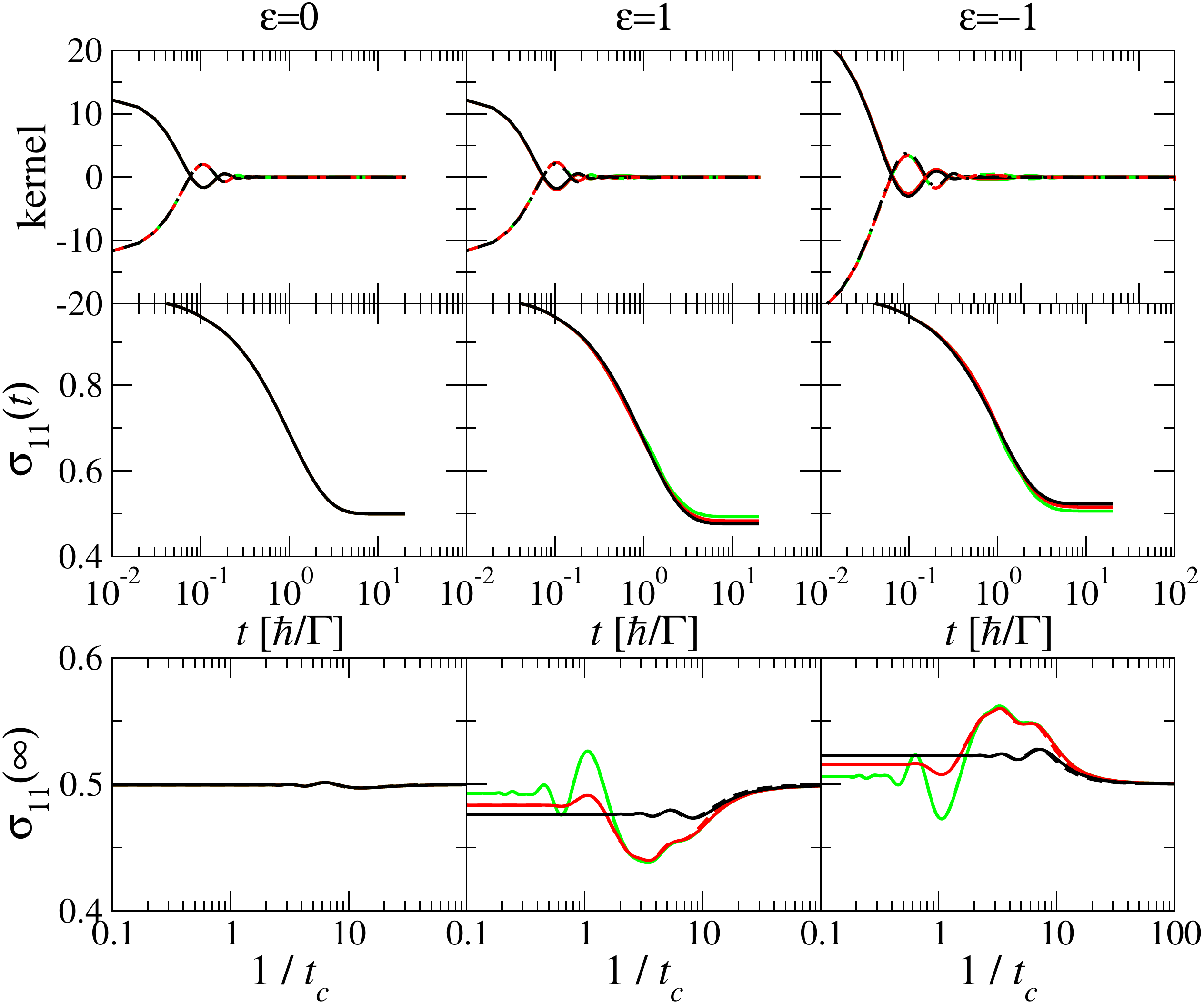}\protect\caption{ \label{fig:V=00003D10. different temps.} Same as Fig\ \ref{fig:V=00003D1 different temps}
but for $V=10\Gamma$.}
\end{figure}

\begin{figure}[t]
\includegraphics[width=8cm]{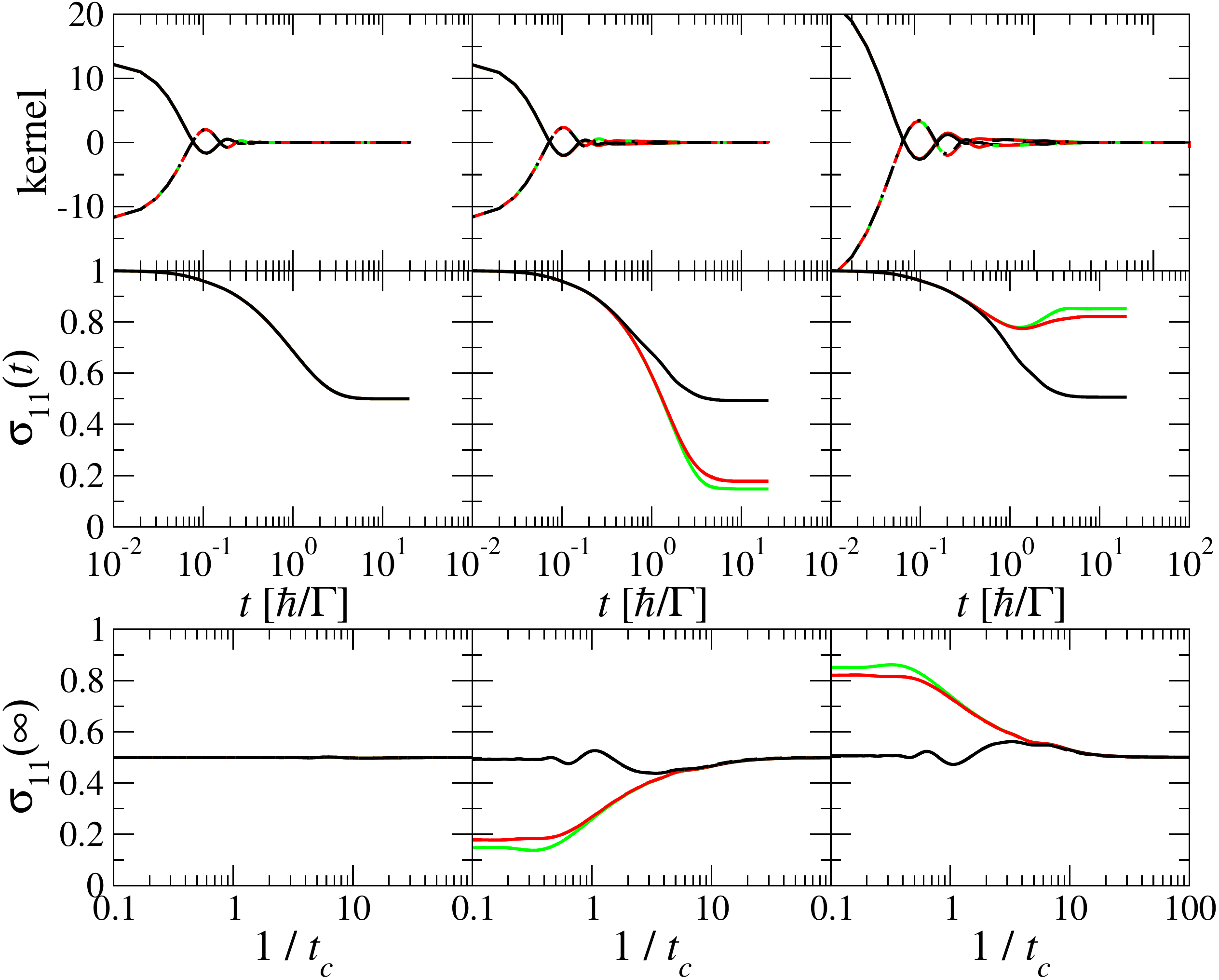}\protect\caption{\label{fig:T=00003D0.1} Same as Fig.\ \ref{fig:V=00003D1 different temps}
but for a fixed temperature $T=\Gamma/10$. Green, red and black curves
correspond to $V_{{\rm SD}}=0$, $V_{{\rm SD}}=\Gamma$, and $V_{{\rm SD}}=10\Gamma$,
respectively. }
\end{figure}

In the upper panels of Figs.~\ref{fig:V=00003D1 different temps},
\ref{fig:V=00003D10. different temps.} and \ref{fig:T=00003D0.1}
we compare the $'00,00'$ element of the TC memory kernel to the time-derivative
of the corresponding TCL element for several values of the dot energy
($\varepsilon$), for three temperatures, and for different values
of the bias-voltage ($V_{{\rm SD}}$). Since the TCL generator decays
to a constant, we plot its time-derivative, providing a more direct
comparison to the TC kernel. $\kappa_{00,00}\left(t\right)$ and ${\cal \partial K}_{00,00}\left(t\right)/\partial t$
have the same sign, and thus for clarity we plot $-\kappa_{00,00}\left(t\right)$,
however, even when we compare the two quantities directly, they show
a slightly different transient behavior. Interestingly, both $\kappa\left(t\right)$
and ${\cal \dot{K}}\left(t\right)$ decay on an identical time scale,
regardless of the dot energy, temperature or source-drain bias voltage. 

In the middle panels of Figs.\ \ref{fig:V=00003D1 different temps},
\ref{fig:V=00003D10. different temps.} and \ref{fig:T=00003D0.1}
we show the transient dot population ($\sigma_{11}\left(t\right)$)
using the TC and TCL approaches. In addition, we also plot the exact
dot population obtained directly from the NEGF calculation. Thus,
each panel includes $9$ curves, but only $3$ are clearly distinguishable,
corresponding to different parameters. The other curves overlap signifying
the excellent agreement between the TC, TCL and the exact NEGF results
for the cutoff time ($t_{c}$) used. Furthermore, it is clearly evident
that the decay of the dot population occurs on longer time scales
compared to the decay of $\kappa_{00,00}\left(t\right)$ or ${\cal \partial K}_{00,00}\left(t\right)/\partial t$.
This is true as long as the source-drain bias voltage is not too large
and the temperature not too small. This separation of time scales
is central to the use of the TC and TCL approaches.

In the lower panels of Figs.\ \ref{fig:V=00003D1 different temps},
\ref{fig:V=00003D10. different temps.} and \ref{fig:T=00003D0.1}
we plot the steady state value of $\sigma_{11}\left(t\rightarrow\infty\right)$
as obtained directly from the kernel elements:
\begin{equation}
\sigma_{11}\left(\infty\right)=\frac{\mathcal{K}_{00}}{\mathcal{K}_{00}-\mathcal{K}_{01}},
\end{equation}
where for the TC ${\cal K}_{ij}\equiv\int_{0}^{t_{c}}dt\kappa_{ii,jj}\left(t\right)$
and for the TCL ${\cal K}_{ij}={\cal K}_{ii,jj}$$\left(t_{c}\right)$.
For all cases studied, the TC and TCL show exactly the same behavior
with the cutoff time, $t_{c}.$ This implies that neither formulation
has an advantage over the other with respect to the time scales needed
to generate the kernel using a proper impurity solver. Whether this
is a general result or specific to the resonant level model is still
an open question, which will be addressed in future studies.

\section{Concluding remarks }

We have adopted the reduced density propagator formalism to obtain
the TCL generator, circumventing the need to invert a super-operator
in the full Hilbert space and thereby use a perturbative scheme in
the system-bath coupling. The elements of the reduced density propagator
can be deduced from the time-dependent reduced density matrix alone,
which can be generated using a proper impurity solver or a diagrammatic
technique. The formalism provides a clear advantage for situations
where the TCL generator decays much faster than the reduced density
matrix itself.

We have implemented the approach for an open non-interacting quantum
system driven away from equilibrium by a source-drain bias potential.
Comparison of the present TCL approach to our previously developed
TC approach reveals that the relaxation time for both kernels is identical,
suggesting that neither one is superior in this respect. However,
the current formalism requires only system observables to generate
the kernel. This may become an advantage, depending on the specific
choice of the impurity solver used to calculate the kernel.

\begin{acknowledgments}
We would like to thank David Reichman and Michael Thoss for insightful
discussions and suggestions. E.Y.W. is grateful to The Center for
Nanoscience and Nanotechnology at Tel Aviv University for a doctoral
fellowship.
\end{acknowledgments}

\appendix

\section{Tetradic representation of the system propagator\label{apx:Tetradic-representation-of-Usigma}}

To find the tetradic representation for the reduced system propagator,
let us work in the uncoupled base, with Latin letters for the dot
many-body states $i,j,k,l\ldots$ and Greek for the leads many-body
states $\alpha,\beta,\gamma,\delta\ldots$. Then, we may write:
\begin{eqnarray}
\rho\left(t\right) & = & {\cal U}\left(t\right)\rho\left(0\right)\nonumber \\
 & \Downarrow\nonumber \\
\rho_{i\alpha,j\beta}\left(t\right) & = & \sum_{k\gamma,l\delta}{\cal U}_{i\alpha,j\beta;k\gamma,l\delta}\left(t\right)\rho_{k\gamma,l\delta}\left(0\right)\label{eq:rho_i,alpha,j,beta}\\
 & = & \sum_{k\gamma,l\delta}\left(e^{\mathcal{L}t}\right)_{i\alpha,j\beta;k\gamma,l\delta}\left(t\right)\rho_{k\gamma,l\delta}\left(0\right)\nonumber \\
 & = & \sum_{k\gamma,l\delta}\left(e^{iHt}\right)_{i\alpha,k\gamma}\left(e^{-iHt}\right)_{l\delta,j\beta}\rho_{k\gamma,l\delta}\left(0\right)\nonumber \\
 & = & \sum_{k\gamma,l\delta}\left(e^{iHt}\right)_{i\alpha,k\gamma}\left(e^{-iHt}\right)_{l\delta,j\beta}\rho_{B,\gamma\delta}\left(0\right)\sigma_{kl}\left(0\right)\nonumber 
\end{eqnarray}
where we used the explicit tetradic form of the Liouville propagator,
as shown in reference \onlinecite{zwanzig_nonequilibrium_2001} (C.6
P.105): 
\begin{equation}
\left(e^{\mathcal{L}t}\right)_{ij,kl}=\left(e^{iHt}\right)_{ik}\left(e^{-iHt}\right)_{lj}
\end{equation}
and the factorized initial conditions: 
\begin{align}
\rho_{k\gamma,l\delta}\left(0\right) & =\left\langle k,\gamma\left|\rho\left(0\right)\right|l,\delta\right\rangle \nonumber \\
 & =\left\langle k\right|\otimes\left\langle \gamma\right|\left(\rho_{B}\left(0\right)\otimes\sigma\left(0\right)\right)\left|l\right\rangle \otimes\left|\delta\right\rangle \nonumber \\
 & =\left\langle \gamma\left|\rho_{B}\left(0\right)\right|\delta\right\rangle \left\langle k\left|\sigma\left(0\right)\right|l\right\rangle =\rho_{B,\gamma\delta}\left(0\right)\sigma_{kl}\left(0\right).
\end{align}

Now performing the trace over the bath is simply a matter of summing
the expression in the last line of Eq.\ \eqref{eq:rho_i,alpha,j,beta}
over $\alpha$ with $\delta_{\alpha\beta}$, and we have:
\begin{eqnarray}
\sigma_{ij}\left(t\right) & = & \left(\mathrm{Tr}_{B}\left\{ \rho\left(t\right)\right\} \right)_{ij}\nonumber \\
 & = & \sum_{\alpha}\sum_{k\gamma,l\delta}\left(e^{iHt}\right)_{i\alpha,k\gamma}\left(e^{-iHt}\right)_{l\delta,j\alpha}\rho_{B,\gamma\delta}\left(0\right)\sigma_{kl}\left(0\right)\nonumber \\
 & = & \sum_{kl}\left[\sum_{\alpha,\gamma,\delta}\left(e^{iHt}\right)_{i\alpha,k\gamma}\left(e^{-iHt}\right)_{l\delta,j\alpha}\rho_{B,\gamma\delta}\left(0\right)\right]\sigma_{kl}\left(0\right)\nonumber \\
 & \equiv & \sum_{kl}{\cal U}_{S,ij,kl}\left(t\right)\sigma_{kl}\left(0\right)
\end{eqnarray}
where the tetradic form of the system propagator is given by:
\begin{equation}
{\cal U}_{S,ij,kl}\left(t\right)=\sum_{\alpha,\beta,\gamma}\left(e^{iHt}\right)_{i\alpha,k\beta}\rho_{B,\beta\gamma}\left(0\right)\left(e^{-iHt}\right)_{l\gamma,j\alpha}.
\end{equation}

\end{document}